\documentclass[preprint]{elsarticle}

\usepackage{comment}
\usepackage{lineno,hyperref}
\modulolinenumbers[5]

\journal{XYZ}

\usepackage[utf8]{inputenc}
\usepackage[english]{babel}
\usepackage{amsmath}
\usepackage{amsfonts}
\usepackage{amssymb}
\usepackage{graphicx}

\usepackage{hyperref}
\usepackage{amsfonts}
\usepackage{amsmath}
\usepackage{amsthm}
\usepackage{mathrsfs}
\usepackage{amssymb}
\usepackage[numbers]{natbib}
\usepackage[fit]{truncate}
\usepackage{mathrsfs}

\usepackage{latexsym}
\usepackage{amssymb}
\usepackage{graphicx}
\usepackage{graphicx,epsfig}

\usepackage{amsmath}
\usepackage{amsthm}
\usepackage{colortbl}
\usepackage{hyperref}
\usepackage{framed,color}
\definecolor{shadecolor}{rgb}{1,0.8,0.3}
\usepackage{color}
\usepackage{mathtools}
\usepackage{amsthm}
\usepackage{rotating}
\usepackage{subfig}

\usepackage{graphicx}
\usepackage{epstopdf}
\usepackage[utf8]{inputenc}
\usepackage[english]{babel}

\theoremstyle{definition}
\newtheorem{definition}{Definition}[section]
 
 \theoremstyle{Hypothesis}

\newdefinition{rmk}{Remark}
\newproof{pf}{proof}
\newproof{pot}{Proof of Theorem \ref{t1}}
\newproof{pott}{Proof of Theorem \ref{t2}}

\newcommand{\distas}[1]{\mathbin{\overset{#1}{\kern\z@\sim}}}%
\newsavebox{\mybox}\newsavebox{\mysim}
\newcommand{\distras}[1]{%
  \savebox{\mybox}{\hbox{\kern3pt$\scriptstyle#1$\kern3pt}}%
  \savebox{\mysim}{\hbox{$\sim$}}%
  \mathbin{\overset{#1}{\kern\z@\resizebox{\wd\mybox}{\ht\mysim}{$\sim$}}}%
}
\makeatother
\newcommand {\ctn}{\citet} % change to \citet if using natbib
\usepackage{graphicx,amsmath,latexsym,amssymb}
\usepackage{float,epsfig,multirow,rotating,times}
\usepackage{upgreek,wrapfig}
\usepackage{comment}
\usepackage{slashbox}

\newcommand{\bL}{\boldsymbol{L}}

\newcommand{\btheta}{\boldsymbol{\theta}}

\newcommand{\bbeta}{\boldsymbol{\beta}}

\newcommand{\bxi}{\boldsymbol{\xi}}
\newcommand{\bGamma}{\boldsymbol{\Gamma}}

\newcommand{\bSigma}{\boldsymbol{\Sigma}}

\newcommand{\bepsilon}{\boldsymbol{\epsilon}}

\newcommand{\bD}{\boldsymbol{D}}

\newcommand{\bV}{\boldsymbol{V}}

\newcommand{\bh}{\boldsymbol{h}}
\newcommand{\bH}{\boldsymbol{H}}

\newcommand{\bI}{\boldsymbol{I}}

\newcommand{\bK}{\boldsymbol{K}}

\newcommand{\bS}{\boldsymbol{S}}

\newcommand{\bu}{\boldsymbol{u}}

\newcommand{\bs}{\boldsymbol{s}}

\newcommand{\bx}{\bm{x}}

\newcommand{\by}{\boldsymbol{y}}
\newcommand{\bY}{\boldsymbol{Y}}

\newcommand{\bzero}{\boldsymbol{0}}

\newtheorem{theorem}{Theorem}

\newcommand{\e}{\ensuremath{\epsilon}}

\newcommand{\be}{\boldsymbol\e}
\newcommand{\bm}{\mathbf}

\numberwithin{equation}{section}
\numberwithin{algo}{section}
\numberwithin{table}{section}
\numberwithin{figure}{section}

\begin{document}

\begin{frontmatter}

\title{A Statistical Perspective on Inverse and Inverse Regression Problems}%\tnoteref{mytitlenote}}

%\author{Debashis Chatterjee \fnref{myfootnote}}

\author[mymainaddress]{Debashis Chatterjee}
%\corref{mycorrespondingauthor}}
%\cortext[mycorrespondingauthor]{Corresponding author. This is part of the Ph.D. research work of the first author 
%which is in progress  at the Indian Statistical Institute}
\ead{debashis1chatterjee@gmail.com}

\author[mymainaddress]{Sourabh Bhattacharya\corref{mycorrespondingauthor}}
\cortext[mycorrespondingauthor]{Corresponding author.} 
\ead{bhsourabh@gmail.com, sourabh@isical.ac.in}

\address[mymainaddress]{Interdisciplinary Statistical Research Unit, Indian Statistical Institute,
\newline 203 B. T. Road, Kolkata - 700108, India}

\begin{abstract}
Inverse problems, where in broad sense the task is to learn from the noisy response about some unknown function, 
usually represented as the argument of some known functional form, has received wide attention
in the general scientific disciplines. However, in mainstream statistics such inverse problem paradigm
does not seem to be as popular. In this article we provide a brief overview of such problems from
a statistical, particularly Bayesian, perspective. 

We also compare and contrast the above class of problems with the perhaps more statistically familiar
inverse regression problems, arguing that this class of problems contains the traditional class of inverse problems.
In course of our review we point out that the statistical literature is very scarce with respect to both the inverse paradigms,
and substantial research work is still necessary to develop the fields. 

\end{abstract}


\end{frontmatter}

%\linenumbers

\section{Introduction}
\label{S1}

The similarities and dissimilarities between inverse problems and the more traditional forward problems are usually
not clearly explained in the literature, and often ``ill-posed" is the term used to loosely characterize inverse problems.
We point out that these two problems may have the same goal or different goal, while both consider the same
model given the data. We first elucidate using the traditional case
of deterministic differential equations, that the goals of the two problems may be the same. Consider a dynamical system 
\begin{equation}
\frac{dx_t}{dt}=G(t,x_t,\theta), 
\label{eq:diff1}
\end{equation}
where $G$ is a known function and $\theta$ is a parameter. In the forward problem the goal
is to obtain the solution $x_t\equiv x_t(\theta)$, given $\theta$ and the initial conditions, whereas, in the inverse problem, 
the aim is to obtain $\theta$ given the solution process $x_t$. 
Realistically, the differential equation would be perturbed
by noise, and so, one observes the data $\by=(y_1,\ldots,y_T)^T$, 
where 
\begin{equation}
y_t=x_t(\theta)+\epsilon_t, 
\label{eq:model1}
\end{equation}
for noise variables $\epsilon_t$ having
some suitable independent and identical ($iid$) error distribution $q$, which we assume to be known for simplicity
of illustration. 
A typical method of estimating $\theta$, employed by the scientific community, is the method of calibration, 
where the solution of (\ref{eq:diff1}) would be obtained for each $\theta$-value on a proposed grid of 
plausible values, and a set $\tilde\by(\theta)=(\tilde y_1(\theta),\ldots,\tilde y_T(\theta))^T$ 
is generated from the model (\ref{eq:model1})
for every such $\theta$ after simulating, for $i=1,\ldots,T$, $\tilde\epsilon_t\stackrel{iid}{\sim}q$;
then forming $\tilde y_t(\theta)=x_t(\theta)+\tilde\epsilon_t$, and finally reporting that value 
$\theta$ in the grid as an estimate of the true values for which $\|\by-\tilde\by(\theta)\|$ is minimized, 
given some distance measure $\|\cdot\|$; maximization of the correlation between $\by$ and $\tilde\by(\theta)$
is also considered. In other words, the calibration method makes use of the forward technique
to estimate the desired quantities of the model. On the other hand, the inverse problem paradigm attempts to
directly estimate $\theta$ from the observed data $\by$ usually by minimizing some discrepancy
measure between $\by$ and $\bx(\theta)$, where $\bx(\theta)=(x_1(\theta),\ldots,x_T(\theta))^T$. 
Hence, from this perspective the goals of both forward and inverse approaches are the same, that is,
estimation of $\theta$. However, the forward approach is well-posed, whereas, the inverse approach is often
ill-posed. To clarify, note that within a grid, there always exists some $\hat\theta$ that minimizes 
$\|\by-\tilde\by(\theta)\|$ among all the grid-values. In this sense the forward problem may be thought of as well-posed. 
However, direct minimization of the discrepancy between $\by$ and $\bx(\theta)$ with respect to $\theta$ is usually difficult
and for high-dimensional $\theta$, the solution to the minimization problem is usually not unique, and small perturbations
of the data causes large changes in the possible set of solutions, so that the inverse approach is usually ill-posed.
Of course, if the minimization is sought over a set of grid values of $\theta$ only, then the inverse problem
becomes well-posed.

From the statistical perspective, the unknown parameter $\theta$ of the model needs to be learned, in
either classical or Bayesian way, and hence, in this sense there is no real distinction between forward and inverse problems. 
Indeed, statistically, since the data are modeled conditionally on the parameters, all problems where learning 
the model parameter given the data is the goal, are inverse problems.
We remark that the literature usually considers learning unknown functions from the data in the realm of inverse
problems, but a function is nothing but an infinite-dimensional parameter, which is a very common learning problem in statistics.

We now explain when forward and inverse problems can differ in their aims, and are significantly different even
from the statistical perspective. 
To give an example, consider the palaeoclimate reconstruction problem discussed in \ctn{Haslett06} where 
the reconstruction of prehistoric climate at Glendalough in Ireland from fossil pollen is of interest. The 
model is built on the realistic assumption that pollen abundance depends upon climate, not the other way around.
The  compositional pollen data with the modern climates are available at many modern sites but the climate values 
associated with the fossil pollen data are missing. 
The inverse nature of the problem is associated with the fact that it is of interest to 
predict the fossil climate values, given the pollen assemblages. 
The forward problem would result, if given the fossil climate values (if known), the fossil pollen abundances (if unknown),
were to be predicted. 

Technically, 
given a data set  $\by$ that depends upon covariates $\bx$, with a probability distribution $f(\by | \bx, \theta )$ 
where $\theta$ is the model parameter, we call the problem `inverse' if it is of interest to predict the 
corresponding unknown $\tilde x$ given a new observed $\tilde y$ (see \ctn{Bhattacharya07}), 
after eliminating $\theta$. On the other hand, 
the more conventional forward problem considers the prediction of $\tilde y$ for given $\tilde x$ with the same 
probability distribution, again, after eliminating the unknown parameter $\theta$. 
This perspective clearly distinguishes the forward and inverse problems, as opposed to the other parameter-learning 
perspective discussed above, which is much more widely considered in the literature. In fact, with respect to
predicting unknown covariates from the responses, mostly inverse
linear regression, particularly in the classical set-up, has been considered in the literature.
To distinguish the traditional inverse problems from the covariate-prediction perspective, 
we use the phrase `inverse regression' to refer to the latter. Other examples of inverse regression are given in
Section \ref{sec:inv_reg}.

Our discussion shows that statistically, there is nothing special about the existing literature on inverse 
problems that considers
estimation of unknown (perhaps, infinite-dimensional) parameters, and the only class of problems that can be
truly regarded as inverse problems as distinguished from forward problems are those which consider prediction
of unknown covariates from the dependent response data. However, for the sake of completeness, the traditional 
inverse problems related to learning of unknown functions shall occupy a 
significant portion of our review.

%Inverse problems are becoming most of theoretical and of practical curiosity, where the quantity of interest cannot be 
%observed directly from the experiment and can be treated as covariates of the observed erroneous response. 
%There are very few Bayesian literature available for addressing  model consistency and model selection problem in 
%inverse problems. Recently, in \ctn{rs1}, \ctn{vol} and \ctn{the} the authors have addressed the inverse model 
%representation and addressed the posterior consistency issue for inverse problem under certain assumptions. 
%In \ctn{al1} inverse model selection problem has been discussed for very particular class of problems.

The rest of the paper is structured as follows. In Section \ref{sec:trad_inv_prob} we discuss the general inverse model,
providing several examples. In Section \ref{sec:linear_inverse_problem} we focus on linear inverse problems, which
constitute the most popular class of inverse problems, and review the links between the Bayesian approach based
on simple finite difference priors and the deterministic Tikhonov regularization.
Connections between Gaussian process based Bayesian inverse problems and deterministic regularizations
are reviewed in Section \ref{sec:link_GP_tik}.
In Section \ref{sec:diff_gp} we provide an overview of the connections between 
the Gaussian process based Bayesian approach and regularization using differential operators, which generalizes
the discussion of Section \ref{sec:linear_inverse_problem} on the connection between finite difference priors 
and the Tikhonov regularization. The Bayesian approach to inverse problems in Hilbert spaces is discussed in
Section \ref{sec:Bayesian_Hilbert}.
We then turn attention to inverse regression problems, providing an overview of such problems and discussing 
the links with traditional inverse problems in Section \ref{sec:inv_reg}. Finally, we make concluding remarks
in Section \ref{sec:conclusion}.

\section{Traditional inverse problem}
\label{sec:trad_inv_prob}
%Let $\mathbb{A}$ and $\mathbb{B}$ be Hilbert spaces. 
Suppose that one is interested in learning about 
the function $\theta$ given the noisy observed responses %$\by=\{y_1,\ldots,y_n\}$, with 
$\by_n=(y_1,\ldots,y_n)^T$, where the relationship between 
$\theta$ and $\by_n$ is governed by following equation ~\eqref{e1} :
\begin{equation}
\label{e1}
y_i= G(x_i,\theta) + \epsilon_i,
\end{equation}  
for $i=1,\ldots,n$,
where $x_i$ are known covariates or design points, $\epsilon_i$ are errors associated with the $i$-th observation 
and $G$ is a forward operator defined appropriately, which  
%$G : \mathbb{A} \rightarrow \mathbb{B}$ is the forward 
%operator from $\mathbb{A}$ to $\mathbb{B}$.
%$\epsilon_i \in \mathbb{B}$ is the error associated with the $i$-th observation. 
is usually allowed to be non-injective. 
%Moreover, standard approximate inversion of $G$ may not serve the purpose as except some statistical properties, 

Note that since $\bepsilon_n=(\epsilon_1,\ldots,\epsilon_n)^T$ is unknown, the noisy observation vector $\by_n$ 
itself may not be in the image set of $G$. If $\theta$ is a $p$-dimensional parameter, then there will often be 
situations when the number of equations is smaller than the number of unknowns, in the sense
that $p>n$ (see, for example, \ctn{Dashti15}).  
Modern statistical research is increasingly coming across such inverse problems 
termed  as ``ill-posed" which are not in the exact domain of statistical estimation procedures (\ctn{Finbarr86a}) where the 
maximum likelihood solution or classical least squares may not be uniquely defined and with very bad perturbation 
sensitivity of the classical solution. However, although such problematic issues are said to characterize
inverse problems, the problems in fact fall in the so-called ``large $p$ small $n$" paradigm and has received
wide attention in statistics; see, for example, \ctn{Buhlmann11}, \ctn{Giraud15}. A key concept involved in handling
such problems is inclusion of some appropriate penalty term in the discrepancy to be minimized with respect to $\theta$.
Such regularization methods are initiated by \ctn{Tikhonov63} and \ctn{Tikhonov77}. Under this method, usually a criterion
of the following form is chosen for the minimization purpose:
\begin{equation}
\frac{1}{n}\sum_{i=1}^n\left[y_i-G(x_i,\theta)\right]^2+\lambda J(\theta),~\lambda>0.
\label{eq:tik1}
\end{equation}
The functional $J$ is chosen such that highly implausible or irregular values of $\theta$ has large values (\ctn{Finbarr86a}).
Thus, depending on the problem at hand, $J(\theta)$ can be used to induce ``sparsity" in an appropriate sense so 
that the minimization problem may be well-defined.
We next present several examples of classical inverse problems based on \ctn{Aster13}.

%Special cases of this general formulation will be illustrated subsequently.
 %Let $\Vert \cdot \Vert_{\mathbb{B}}$ be the least square norm of  $\mathbb{B}$. The first approach could be to find $\displaystyle \text{argmin}_{x \in \mathbb{A}} \Vert y-G(x)\Vert^{2}_{\mathbb{B}} $. Assuming the iterative solution points $x^{(n)}$ converges. But the assumption is often not true. To overcome the problem, the next approach that can be taken is to regularize by taking a suitably chosen point $m_{0} \in E$ for $E \subset \mathbb{A}$ and then find solution of \eqref{regu} for suitably chosen  positive reals $\delta_1$ and $\delta_2$ with $\delta_1 +\delta_2=1$.
%\begin{equation}
%\label{regu}
% \displaystyle \text{argmin}_{x \in E} \left( \delta_{1} \Vert y-G(x) \Vert^{2}_{\mathbb{B}} +  \delta_{2} \Vert x-m_{0}  \Vert^{2}_{E} \right).
% \end{equation}
%  The above problem is known as regularized minimization problem, details of which can be found in \ctn{rs1}.

\subsection{Examples of inverse problems}
\label{subsec:examples_inverse_problems}
\subsubsection{Vertical seismic profiling}
\label{subsubsec:inv_example1}
In this scientific field, one wishes to learn about the vertical seismic velocity of the
material surrounding a borehole. A source generates downward-propagating seismic wavefront at the surface, 
and in the borehole, a string of seismometers sense these seismic waves.
The arrival times of the seismic wavefront at each instrument are measured from the recorded seismograms.
These times provide information on the seismic velocity for vertically traveling waves as a function of depth.
The problem is nonlinear if it is expressed in terms of seismic
velocities. However, we can linearize this problem via a simple change of variables, as follows.
Letting $z$ denote the depth, it is possible to parameterize the seismic structure in terms of slowness, 
$s(z)$, which is the reciprocal of the velocity $v(z)$. The observed travel time at depth $z$ can then be 
expressed as: 
%the definite
%integral of the vertical slowness, s, from the surface to z:
\begin{equation}
t(z)=\int_0^zs(u)du=\int_0^{\infty}s(u)H(z-u)du,
\label{eq:heaviside}
\end{equation}
where $H$ is the Heaviside step function. The interest is to learn about $s(z)$ given observed $t(z)$.
Theoretically, $s(z)=\frac{dt(z)}{dz}$, but in practice, simply differentiating the observations need not lead to
useful solutions because noise is generally present in the observed times $t(z)$, and naive differentiation
may lead to unrealistic features of the solution.

\subsubsection{Estimation of buried line mass density from vertical gravity anomaly}
\label{subsubsec:inv_example2}
Here the problem is to estimate an unknown buried line mass density $m(x)$ from data on 
vertical gravity anomaly,
$d(x)$, observed at some height, $h$.
The mathematical relationship between $d(x)$ and $m(x)$ is given by
\begin{equation*}
d(x)=\int_{-\infty}^{\infty}\frac{h}{\left[(u-x)^2+h^2\right]^{\frac{3}{2}}}m(u)du.
\end{equation*}
As before, noise in the data renders the above linear inverse problem difficult.
Variations of the above example has been considered in \ctn{Aster13}.

\subsubsection{Estimation of incident light intensity from diffracted light intensity}
\label{subsubsec:inv_example3}
Consider an experiment in which an angular distribution
of illumination passes through a thin slit and produces a diffraction pattern, for which
the intensity is observed.
The data, $d(s)$, are measurements of diffracted light intensity as a function of the
outgoing angle $-\pi/2\leq s\leq\pi/2$. The goal here is to obtain the intensity of incident light
on the slit, $m(\theta)$, as a function of the incoming angle $-\pi/2\leq\theta\leq \pi/2$, using the
following mathematical relationship:
\begin{equation*}
d(s)=\int_{-\pi/2}^{\pi/2}\left(\cos(s)+\cos(\theta)\right)^2
\left(\frac{\sin(\pi\left(\sin(s)+\sin(\theta)\right))}{\pi\left(\sin(s)+\sin(\theta)\right)}\right)^2m(\theta)d\theta.
\end{equation*}

\subsubsection{Groundwater pollution source history reconstruction problem}
\label{subsubsec:inv_example4}
Consider the problem of recovering the history of groundwater pollution at a source
site from later measurements of the contamination at downstream wells to which the
contaminant plume has been transported by advection and diffusion. The mathematical model
for contamination transport is given by the following advection-diffusion equation with respect to $t$ and transported site
$x$:
\begin{align}
\frac{\partial C}{\partial t}&=D\frac{\partial^2 C}{\partial x^2}-\nu\frac{\partial C}{\partial x}\notag\\
C(0,t) &= C_{in}(t)\notag\\
C(x,t) &\rightarrow 0~\mbox{as}~x\rightarrow\infty.\notag
\end{align}
In the above, $D$ is the diffusion coefficient, $\nu$ is the velocity of the groundwater flow,
and $C_{in}(t)$ is the time history of contaminant injection at $x = 0$. The solution
to the above advection-diffusion equation is given by
\begin{equation*}
C(x,T)=\int_0^TC_{in}(t)f(x,T-t)dt,
\end{equation*}
where
\begin{equation*}
f(x,T-t)=\frac{x}{2\sqrt{\pi D(T-t)^3}}\exp\left[\frac{\left(x-\nu(T-t)\right)^2}{4D(T-t)}\right].
\end{equation*}
It is of interest to learn about $C_{in}(t)$ from data observed on $C(x,T)$.

\subsubsection{Transmission tomography}
\label{subsubsec:inv_example5}
%The physical model for tomography in its most basic form (Figure 1.8) assumes that
%geometric ray theory (essentially the high-frequency limiting case of the wave equation)
%is valid, so 
The most basic physical model for tomography assumes that wave energy traveling between 
a source and receiver can be considered to
be propagating along infinitesimally narrow ray paths. 
%The density of ray path coverage
%in a tomographic problem may vary significantly throughout a section or volume, and
%provide much better constraints on physical properties in densely sampled regions than
%in sparsely sampled ones.
In seismic tomography, if the slowness at a point $x$ is $s(x)$, and the ray path is
known, then the travel time for seismic energy transiting along that ray path is given by
the line integral along $\ell$:
\begin{equation}
t=\int_{\ell}s(x(l))dl.
\label{eq:tt}
\end{equation}
Learning of $s(x)$ from $t$ is required.
Note that (\ref{eq:tt}) is a high-dimensional generalization of (\ref{eq:heaviside}).
In reality, seismic ray paths will be
bent due to refraction and/or reflection, resulting in nonlinear inverse problem. 
%In cases where such effects are negligible, ray
%paths can be usefully approximated as straight lines (e.g., as depicted in Figure 1.8),
%and the forward and inverse problems can be cast in a linear form. However, if the ray
%paths are significantly bent by slowness variations, the resulting inverse problem will be
%nonlinear

The above examples demonstrate the ubiquity of linear inverse problems. As a result, 
in the next section we take up the case of linear inverse problems and illustrate the Bayesian approach in details,
also investigating connections with the deterministic approach employed by the general scientific community. 

\section{Linear inverse problem}
\label{sec:linear_inverse_problem}

The motivating examples and discussions in this section are based on \ctn{Bui12}.

Let us consider the following one-dimensional integral equation on a finite interval as in equation \eqref{l1}:
\begin{equation}
\label{l1}
G(x,\theta)=\int K(x,t)\  \theta(t)  \ dt,
\end{equation} 
where $K(x,\cdot)$ is some appropriate, known, real-valued function given $x$ 
%$K : \mathbf{D}(K) \subset \mathbb{A} \rightarrow \mathbb{B}$ is an operator defined on some subspace 
%$\mathbf{D}(K)$ of a normed linear space $\mathbb{A}$ which has image values in $\mathbb{B}$. 
%We can also represent it in other form like $ K f = g $. 
Now, let the dataset be $\by_n=(y_1, y_2, \ldots, y_n)^T$. Then for a known system response 
$K(x_{i}, t)$ for the dataset, the equation can be written as follows:
\begin{equation}
\label{l2}
y_i=\int G(x_{i},\theta) + \epsilon_i \ ; \ \ \ i\in \{1, 2, \ldots, n\}
\end{equation} 
%where $\bepsilon =\{\epsilon_1, \epsilon_2, \ldots, \epsilon_n \}$ is the vector. 

As a particular example, let $G(x,\theta)=\int_0^1 K(x,t)\  \theta(t)  \ dt$, where 
$K(x,t)=\frac{1}{\sqrt{2\pi\psi^2}}\exp\left\{-(x-t)^2/2\psi^2\right\}$ is the Gaussian kernel
and $\theta:[0,1]\mapsto\mathbb R$ is to be learned given the data $\by_n$ and $\bx_n=(x_1,\ldots,x_n)^T$.
We first illustrate the Bayesian approach and draw connections with the traditional approach of Tikhonov's
regularization when the integral in $G$ is discretized. In this regard, let $x_i=(i-1)/n$, for $i=1,\ldots,n$.
Letting $\btheta=(\theta(x_1),\ldots,\theta(x_n))^T$ and $\bK$ be the $n\times n$ matrix with the 
$(i,j)$-th element $K(x_i,x_j)/n$,
and $\bepsilon_n=(\epsilon_1,\ldots,\epsilon_n)^T$,
the discretized version of (\ref{l2}) can be represented as
\begin{equation}
\by_n=\bK\btheta+\bepsilon_n.
\label{eq:inv1}
\end{equation}
We assume that $\bepsilon_n\sim N_n\left(\bzero_n,\sigma^2\bI_n\right)$, that is, an $n$-variate normal
with mean $\bzero_n$, an $n$-dimensional vector with all components zero, and covariance $\sigma^2\bI_n$,
where $\bI_n$ is the $n$-th order identity matrix.

\subsection{Smooth prior on $\theta$}
\label{subsec:smooth_prior}

To reflect the belief that the function $\theta$ is smooth, one may presume that 
\begin{equation}
\theta(x_i)=\frac{\theta(x_{i-1})+\theta(x_{i+1})}{2}+\tilde\epsilon_i, 
\label{eq:lap1}
\end{equation}
where, for $i=1,\ldots,n$,
$\tilde\epsilon_i\stackrel{iid}{\sim}N\left(0,\tilde\sigma^2\right)$. Thus, {\it a priori}, $\theta(x_i)$
is assumed to be an average of its nearest neighbors to quantify smoothness, with an additive random
perturbation term. Letting 
\begin{equation}
\bL=\frac{1}{2}\left(\begin{array}{cccccc} -1 & 2 & -1 & 0 & \cdots &\cdots\\
                                 0 & -1 & 2 & -1 & 0 & \cdots\\
				 \vdots & \vdots & \vdots & \vdots & \vdots & \vdots\\ 
				 \vdots & \vdots & \vdots & \vdots & \vdots & \vdots\\ 
				   0    &   0    & \cdots & -1  & 2 & -1
				   \end{array}\right),
\label{eq:L}
\end{equation}
and $\tilde\bepsilon=(\tilde\epsilon_1,\ldots,\tilde\epsilon_n)^T$, it follows from (\ref{eq:lap1}) that 				   
\begin{equation}
\bL\btheta=\tilde\bepsilon,
\label{eq:lap2}
\end{equation}
Now, noting that the Laplacian of a twice-differentiable real-valued function $f$ with independent arguments $z_1,\ldots,z_k$
is given by $\Delta f=\sum_{i=1}^k\frac{\partial^2 f}{\partial z^2_i}$, we have 
\begin{equation}
\Delta\theta(x_j)\approx n^2(\bL\btheta)_j, 
\label{eq:approx_lap1}
\end{equation}
where $(\bL\btheta)_j$ is the $j$-th element of $\bL\btheta$. 

However, the rank of $\bL$ is $n-1$, and boundary conditions on the Laplacian operator is necessary to ensure
positive definiteness of the operator. In our case, we assume that $\theta\equiv 0$ outside $[0,1]$, so that
we now assume $\theta(0)=\frac{\theta(x_1)}{2}+\tilde\epsilon_0$ and $\theta(x_n)=\frac{\theta(x_{n-1})}{2}+\tilde\epsilon_n$,
where $\tilde\epsilon_0$ and $\tilde\epsilon_n$ are $iid$ $N\left(0,\tilde\sigma^2\right)$.
With this modification, the prior on $\btheta$ is given by
\begin{equation}
\pi(\btheta)\propto\exp\left(-\frac{1}{2\tilde\sigma^2}\|\tilde\bL\btheta\|^2\right),
\label{eq:prior1}
\end{equation}
where $\|\cdot\|$ is the Euclidean norm and 
\begin{equation}
\tilde\bL=\frac{1}{2}\left(\begin{array}{cccccc} 
                                 2 & -1 & 0 & 0 & \cdots & \cdots\\
				-1 & 2 & -1 & 0 & \cdots &\cdots\\
                                 0 & -1 & 2 & -1 & 0 & \cdots\\
				 \vdots & \vdots & \vdots & \vdots & \vdots & \vdots\\ 
				 \vdots & \vdots & \vdots & \vdots & \vdots & \vdots\\ 
				   0    &   0    & \cdots & -1  & 2 & -1\\
				   0    &   0    & \cdots &  0  & -1 & 2
				   \end{array}\right).
\label{eq:tilde_L}
\end{equation}

Rather than assuming zero boundary conditions, more generally one may assume that $\theta(0)$ and $\theta(x_n)$
are distributed as $N\left(0,\frac{\tilde\sigma^2}{\delta^2_0}\right)$ and $N\left(0,\frac{\tilde\sigma^2}{\delta^2_n}\right)$,
respectively. The resulting modified matrix is then given by
\begin{equation}
\hat\bL=\frac{1}{2}\left(\begin{array}{cccccc} 
                                 2\delta_0 & 0 & 0 & 0 & \cdots & \cdots\\
				-1 & 2 & -1 & 0 & \cdots &\cdots\\
                                 0 & -1 & 2 & -1 & 0 & \cdots\\
				 \vdots & \vdots & \vdots & \vdots & \vdots & \vdots\\ 
				 \vdots & \vdots & \vdots & \vdots & \vdots & \vdots\\ 
				   0    &   0    & \cdots & -1  & 2 & -1\\
				   0    &   0    & \cdots &  0  & 0 & 2\delta_n
				   \end{array}\right).
\label{eq:hat_L}
\end{equation}

To choose $\delta_0$ and $\delta_n$, one may assume that $$Var\left[\theta(0)\right]=\frac{\tilde\sigma^2}{\delta^2_0}
=Var\left[\theta(x_n)\right]=\frac{\tilde\sigma^2}{\delta^2_n}=Var\left[\theta(x_{[n/2]})\right]
=\tilde\sigma^2\be^T_{[n/2]}\left(\hat\bL^T\hat\bL\right)^{-1}\be_{[n/2]},$$
where $[n/2]$ is the largest integer not exceeding $n/2$, and $\be_{[n/2]}$ is the $[n/2]$-th canonical basis
vector in $\mathbb R^{n+1}$. It follows that
$$\delta^2_0=\delta^2_n=\frac{1}{\be^T_{[n/2]}\left(\hat\bL^T\hat\bL\right)^{-1}\be_{[n/2]}}.$$
Since this requires solving a non-linear equation (since $\hat\bL$ contains $\delta_0$ and $\delta_n$), for 
avoiding computational complexity one may simply employ the approximation 
$$\delta^2_0=\delta^2_n=\frac{1}{\be^T_{[n/2]}\left(\tilde\bL^T\tilde\bL\right)^{-1}\be_{[n/2]}},$$
where $\tilde\bL$ is given by (\ref{eq:tilde_L}).

\subsection{Non-smooth prior on $\theta$}
\label{subsec:non_smooth_prior}

To begin with, let us assume that $\theta$ has several points of discontinuities on the grid of points $\{x_0,\ldots,x_n\}$.
To reflect this information in the prior, one may assume that $\theta(0)=0$ and for
$i=1,\ldots,n$, $\theta(x_i)=\theta(x_{i-1})+\tilde\epsilon_i$,
where, as before, $\tilde\epsilon_i$ are $iid$ $N\left(0,\tilde\sigma^2\right)$. 
Then, with
\begin{equation}
\bL^*=\frac{1}{2}\left(\begin{array}{cccccc} 
                                 1 & 0 & 0 & 0 & \cdots & \cdots\\
				-1 & 1 & 0 & 0 & \cdots &\cdots\\
                                 0 & -1 & 1 & 0 & 0 & \cdots\\
				 \vdots & \vdots & \vdots & \vdots & \vdots & \vdots\\ 
				 \vdots & \vdots & \vdots & \vdots & \vdots & \vdots\\ 
				   0    &   0    & \cdots & -1  & 1 & 0\\
				   0    &   0    & \cdots &  0  & - & 1
				   \end{array}\right),
\label{eq:L_star}
\end{equation}
the prior is given by
\begin{equation}
\pi(\btheta)\propto\exp\left(-\frac{1}{2\tilde\sigma^2}\|\bL^*\btheta\|^2\right).
\label{eq:prior2}
\end{equation}
One may also flexibly account for any particular big jump. For instance, if for some $\ell\in\{0,\ldots,n\}$, 
the jump $\theta(x_{\ell})-\theta(x_{\ell-1})$ is particularly large compared to the other jumps, then
it can be assumed that $\theta(x_{\ell})=\theta(x_{\ell-1})+\epsilon^*_{\ell}$, with 
$\epsilon^*_{\ell}\sim N\left(0,\frac{\tilde\sigma^2}{\xi^2}\right)$, where $\xi<1$. Letting $\bD_{\ell}$
be the diagonal matrix with $\xi^2$ being the $\ell$-th diagonal element and $1$ being the other diagonal elements,
the prior is then given by
\begin{equation}
\pi(\btheta)\propto\exp\left(-\frac{1}{2\tilde\sigma^2}\|\bD_{\ell}\bL^*\btheta\|^2\right).
\label{eq:prior3}
\end{equation}

A more general prior can be envisaged where the number and location of the jump discontinuities are unknown. Then
we may consider a diagonal matrix $\bD=diag\{\xi_1,\ldots,\xi_n\}$, so that conditionally on the hyperparameters 
$\xi_1,\ldots,\xi_n$,
the prior on $\btheta$ is given by
\begin{equation}
\pi(\btheta|\xi_1,\ldots,\xi_n)\propto\exp\left(-\frac{1}{2\tilde\sigma^2}\|\bD\bL^*\btheta\|^2\right).
\label{eq:prior4}
\end{equation}
Prior on $\xi_1,\ldots,\xi_n$ may be considered to complete the specification. These may also be estimated
by maximizing the marginal likelihood obtained by integrating out $\btheta$, which is known as the
ML-II method; see \ctn{Berger85}. \ctn{Calvetti07} also advocate likelihood based methods.

\subsection{Posterior distribution}
\label{subsec:posterior}

For convenience, let us generically denote the matrices $\bL$, $\tilde\bL$, $\hat\bL$, $\bL^*$, $\bD_{\ell}\bL^*$,
$\bD\bL^*$, by $\bGamma^{-\frac{1}{2}}$. Then it can be easily verified that the posterior of $\theta$ admits
the following generic form:
\begin{equation}
\pi\left(\btheta|\by_n,\bx_n\right)\propto\exp\left\{-\left[\frac{1}{2\sigma^2}\|\by_n-\bK\btheta\|^2
+\frac{1}{2\tilde\sigma^2}\|\bGamma^{-\frac{1}{2}}\btheta\|^2\right]\right\}.
\label{eq:post1}
\end{equation}
Note that the exponent of the posterior is of the form of the Tikhonov functional, which we denote by $T(\btheta)$. 
The maximizer of the posterior, commonly known as the {\it maximum a posteriori} (MAP) estimator, is given by
\begin{equation}
\hat\btheta_{MAP}=\underset{\btheta}{\arg\max}~\pi\left(\btheta|\by_n,\bx_n\right)=\underset{\btheta}{\arg\min}~T(\btheta).
\label{eq:map1}
\end{equation}
In other words, the deterministic solution to the inverse problem obtained by Tikhonov's regularization is nothing
but the Bayesian MAP estimator in our context.

Writing $\bH=\frac{1}{\sigma^2}\bK^T\bK+\frac{1}{\tilde\sigma^2}\bGamma^{-1}$, which is the Hessian of the
Tikhonov functional (regularized misfit), and writing 
$\|\cdot\|_{\bH}=\|\bH^{\frac{1}{2}}\cdot\|$,
it is clear that (\ref{eq:post1}) can be simplified to the Gaussian form, given by
\begin{equation}
\pi\left(\btheta|\by_n,\bx_n\right)\propto
\exp\left\{-\left\|\btheta-\frac{1}{\sigma^2}\bH^{-1}\bK^{-1}\by_n\right\|^2_{\bH}\right\}.
\label{eq:post2}
\end{equation}
It follows from (\ref{eq:post2}) that the inverse of the Hessian of the regularized misfit is the posterior
covariance itself. From the above posterior it also trivially follows that  
\begin{equation}
\hat\btheta_{MAP}=\frac{1}{\sigma^2}\bH^{-1}\bK^{-1}\by_n
=\frac{1}{\sigma^2}\left(\frac{1}{\sigma^2}\bK^T\bK+\frac{1}{\tilde\sigma^2}\bGamma^{1}\right)^{-1}\bK^T\bY_n,
\label{eq:map2}
\end{equation}
which coincides with the Tikhonov solution for linear inverse problems.
The connection between the traditional deterministic Tikhonov regularization approach with Bayesian analysis
continues to hold even if the likelihood is non-Gaussian.

\subsection{Exploration of the smoothness conditions}
\label{subsec:explore_smoothness}

For deeper investigation of the smoothness conditions, let us write
\begin{equation}
\hat\btheta_{MAP}=\underset{\btheta}{\arg\min}~T(\btheta)=\sigma^2\left(\frac{1}{2}\|\by_n-\tilde\by_n\|^2
+\frac{1}{2}\varrho\|\tilde\bGamma^{\frac{1}{2}}\btheta\|^2\right),
\label{eq:map3}
\end{equation}
where $\tilde\by_n=\bK\btheta$, $\varrho=\sigma^2/\tilde\sigma^2$ and $\tilde\bGamma^{\frac{1}{2}}=\bGamma^{-\frac{1}{2}}$. 
Now, from (\ref{eq:approx_lap1}) it follows that for the smooth priors with the zero boundary conditions, 
our Tikhonov functional discretizes
\begin{equation}
T_{\infty}(\btheta)=\frac{1}{2}\|\by_n-\tilde\by_n\|^2+\frac{1}{2}\varrho\|\Delta\theta\|^2_{L^2(0,1)},
\label{eq:tik_inf1}
\end{equation}
where $\|\cdot\|^2_{L^2(0,1)}=\int_0^1(\cdot)^2dt$. 

On the other hand, for the non-smooth prior (\ref{eq:prior2}), rather than discretizing $\Delta\theta$, $\nabla\theta$, that is,
the gradient of $\theta$, is discretized. In other words, for non-smooth priors, our Tikhonov functional discretizes
\begin{equation}
T_{\infty}(\btheta)=\frac{1}{2}\|\by_n-\tilde\by_n\|^2+\frac{1}{2}\varrho\|\nabla\theta\|^2_{L^2(0,1)}.
\label{eq:tik_inf2}
\end{equation}
Hence, realizations of prior (\ref{eq:prior2}) is less smooth compared to those of our smooth priors. 
However, the realizations (\ref{eq:prior2}) must be continuous.
The priors given by (\ref{eq:prior3}) and (\ref{eq:prior4}) also support continuous functions as long as
the hyperparameters are bounded away from zero. These facts, although clear, can be rigorously justified
by functional analysis arguments, in particular, using the Sobolev imbedding theorem (see, for example,
\ctn{Arbogast08}).

%Let us assume Gaussian prior for the error. 
%In other words, let $\epsilon_i$ are independently and identically distributed with $\epsilon_1 \sim  \mathbf{N}(0, \sigma^2)$. 
%The parameters of the model is then $\theta = ( f, \sigma) \in \Theta$ for some suitable parameter space $\Theta$ 
%with a prior defined on it.  We may assume that $f \sim GP(\mu, R)$ independent of $ \mathbf{\epsilon}$,  where $ GP$ 
%denote Gaussian Process with mean $\mu$ and variance $R$ (for definition of Gaussian process, see Appendix 1 \eqref{gap}).

\section{Links between Bayesian inverse problems based on Gaussian process prior and deterministic regularizations}
\label{sec:link_GP_tik}

In this section, based on \ctn{Rasmussen06}, we illustrate the connections between deterministic 
regularizations such as those obtained from differential operators as above, and Bayesian inverse problems 
based on the very popular Gaussian process prior on the unknown function. A key tool for investigating
such relationship is the reproducing kernel Hilbert space (RKHS).

\subsection{RKHS}
\label{subsec:rkhs}
We adopt the following definition of RKHS provided in \ctn{Rasmussen06}: 
\begin{definition}[RKHS]
\label{def:rkhs}
Let $\mathcal H$ be a Hilbert space
of real functions $\theta$ defined on an index set $\mathfrak X$ . Then $\mathcal H$ is called an RKHS 
endowed with an inner product $\langle\cdot,\cdot\rangle_{\mathcal H}$ (and norm 
$\|\theta\|_{\mathcal H} =\langle \theta,\theta\rangle_{\mathcal H}$) if there exists a function 
$\mathcal K:\mathfrak X\times\mathfrak X\mapsto\mathbb R$ with the following properties:
\begin{itemize}
\item[(a)] for every $x$, $\mathcal K(\cdot,x)\in\mathcal H$, and
\item[(b)] $\mathcal K$ has the reproducing property $\langle \theta(\cdot),\mathcal K(\cdot,x)\rangle_{\mathcal H}
=\theta(x)$.
\end{itemize}
\end{definition}
%See e.g. Sch ̈lkopf and Smola [2002] and Wegman [1982]. 
Observe that since
$\mathcal K(\cdot,x),\mathcal K(\cdot,x')\in\mathcal H$, it follows that 
$\langle \mathcal K(\cdot,x), \mathcal K(\cdot,x')\rangle_{\mathcal  H} = \mathcal K(x,x')$.
The Moore-Aronszajn theorem asserts that the RKHS uniquely determines $\mathcal K$, 
and vice versa. Formally,
\begin{theorem}[\ctn{Aronszajn50}]. 
\label{theorem:moore}
Let $\mathfrak X$ be an index set. 
Then for every positive definite function $\mathcal K(\cdot,\cdot)$ on $\mathfrak X\times\mathfrak X$ there exists
a unique RKHS, and vice versa.
\end{theorem}
Here, by positive definite function $\mathcal K(\cdot,\cdot)$ on $\mathfrak X\times\mathfrak X$, we mean
$\int \mathcal K(x,x')g(x)g(x')d\nu(x)d\nu(x')>0$ for all non-zero functions $g\in L_2\left(\mathfrak X,\nu\right)$,
where $L_2\left(\mathfrak X,\nu\right)$ denotes the space of functions square-integrable on $\mathfrak X$ with respect to
the measure $\nu$.

Indeed, the subspace $\mathcal H_0$ of $\mathcal H$ spanned by the functions 
$\left\{\mathcal K(\cdot,\bx_i);~i=1,2,\ldots\right\}$
is dense in $\mathcal H$ in the sense that every function in $\mathcal H$ is a pointwise limit of a Cauchy sequence
from $\mathcal H_0$.

To proceed, we require the concepts of eigenvalues and eigenfunctions associated with kernels.
In the following section we provide a briefing on these.

\subsection{Eigenvalues and eigenfunctions of kernels}
\label{subsec:eigen}
We borrow the statements of the following definition of eigenvalue and eigenfunction, and the subsequent
statement of Mercer's theorem from \ctn{Rasmussen06}.
\begin{definition}
\label{def:eigen}
A function $\psi(\cdot)$ that obeys the integral equation
\begin{equation}
\int_{\mathfrak X}\mathcal C(x,x')\psi(x) d\nu(x) = \lambda\psi(x'),
\label{eq:eigen}
\end{equation}
is called an eigenfunction of the kernel $\mathcal C$ with eigenvalue $\lambda$ with respect to the 
measure $\nu$. 
\end{definition}

We assume that the ordering is chosen such that $\lambda_1\geq\lambda_2\geq\cdots$.
The eigenfunctions are orthogonal with respect to $\nu$ and can be chosen to be
normalized so that $\int_{\mathfrak X}\psi_i(\bx)\psi_j(\bx)d\nu(x)=\delta_{ij}$,
where $\delta_{ij}=1$ if $i=j$ and $0$ otherwise.

The following well-known theorem (see, for example, \ctn{Konig86}) expresses the positive definite kernel 
$\mathcal C$ in terms of its eigenvalues and eigenfunctions.

\begin{theorem}[Mercer's theorem]
\label{theorem:mercer}
Let $(\mathfrak X,\nu)$ be a finite measure space and $\mathcal C\in L_{\infty}\left(\mathfrak X^2,\nu^2\right)$
be a positive definite kernel. By $L_{\infty}\left(\mathfrak X^2,\nu^2\right)$ we mean
the set of all measurable functions $\mathcal C:\mathfrak X^2\mapsto\mathbb R$ which are essentially bounded,
that is, bounded up to a set of $\nu^2$-measure zero. For any function $\mathcal C$ in this set, its essential supremum,
given by $\inf\left\{C\geq 0:|\mathcal C(x_1,x_2)|<C,~\mbox{for almost all}~(x_1,x_2)\in\mathfrak X\times\mathfrak X\right\}$
serves as the norm $\|\mathcal C\|$.
%such that $T_K : L_2\left(\mathfrak X,\mu\right)\mapsto L_2\left(\mathfrak X,\mu\right)$ is positive
%definite. 

Let $\psi_j\in L_2\left(\mathfrak X,\nu\right)$ be the normalized eigenfunctions of
$\mathcal C$ associated with the eigenvalues $\lambda_j(\mathcal C)>0$. Then
\begin{itemize}
\item[(a)] the eigenvalues $\left\{\lambda_j(\mathcal C)\right\}_{j=1}^{\infty}$ are absolutely summable.
\item[(b)] $\mathcal C(x,x')=\sum_{j=1}^{\infty}\lambda_j(\mathcal C)\psi_j(\bx)\bar{\psi_j}(x')$
holds $\nu^2$-almost everywhere, where the series converges absolutely and uniformly
$\nu^2$-almost everywhere. In the above,
$\bar{\psi_j}$ denotes the complex conjugate of $\psi_j$.
\end{itemize}
\end{theorem}

It is important to note the difference between the eigenvalue $\lambda_j(\mathcal C)$ associated with the kernel 
$\mathcal C$ and $\lambda_j(\bSigma_n)$ where $\bSigma_n$ denotes the $n\times n$ Gram matrix with 
$(i,j)$-th element $\mathcal C(x_i,x_j)$.
Observe that (see \ctn{Rasmussen06}):
\begin{equation}
\lambda_j(\mathcal C)\psi_j(x')=\int_{\mathfrak X}\mathcal C(x,x')\psi_j(x) d\nu(x)
\approx\frac{1}{n}\sum_{i=1}^n\mathcal C(x_i,x')\psi_j(x_i,x'),
\label{eq:eigen_gram}
\end{equation}
where, for $i=1,\ldots,n$, $\bx_i\sim\nu$, assuming that $\nu$ is a probability measure.
Now substituting $x'=x_i$; $i=1,\ldots,n$ in (\ref{eq:eigen_gram}) yields the following
approximate eigen system for the matrix $\bSigma_n$:
\begin{equation}
\bSigma_n\bu_j\approx n\lambda_j(\mathcal C)\bu_j,
\label{eq:eigen_gram2}
\end{equation}
where the $i$-th component of $\bu_j$ is given by 
\begin{equation}
u_{ij}=\frac{\psi_j(x_i)}{\sqrt{n}}. 
\label{eq:u_psi}
\end{equation}
Since $\psi_j$ are
normalized to have unit norm, it holds that
\begin{equation}
\bu^T_j\bu_j=\frac{1}{n}\sum_{i=1}^n\psi^2_j(x_i)\approx\int_{\mathfrak X}\psi^2(x)d\nu(x)=1.
\label{eq:eigen_gram3}
\end{equation}
From (\ref{eq:eigen_gram3}) it follows that 
\begin{equation}
\lambda_j(\bSigma_n)\approx n\lambda_j(\mathcal C). 
\label{eq:eigen_Sigma_K}
\end{equation}
Indeed, Theorem 3.4 of 
\ctn{Baker77} shows that $n^{-1}\lambda_j(\bSigma_n)\rightarrow\lambda_j(\mathcal C)$, as $n\rightarrow\infty$.

For our purposes the main usefulness of the RKHS framework is that $\|\theta\|^2_{\mathcal H}$
can be perceived as a generalization of $\btheta^T\mathcal\bK^{-1}\btheta$, where 
$\btheta=(\theta(x_1),\ldots,\theta(x_n))^T$
and $\mathcal\bK=(\mathcal K(x_i,x_j))_{i,j=1,\ldots,n}$, is the $n\times n$ matrix with $(i,j)$-th element
$\mathcal K(x_i,x_j)$. 

\subsection{Inner product}
\label{subsec:inner_product}
Consider a real positive semidefinite kernel $\mathcal K(x,x')$ with an eigenfunction
expansion $\mathcal K(x, x') =\sum_{i=1}^N\lambda_i\phi_i(x)\phi_i(x')$ relative to a measure $\mu$. 
Mercer's theorem ensures that the eigenfunctions are orthonormal with respect to $\mu$, that is, we have
$\int \phi_i(x)\phi_j(x)d\mu(x) = \delta_{ij}$. Consider a Hilbert space of linear
combinations of the eigenfunctions, that is, $\theta(x) = \sum_{i=1}^N\theta_i\phi_i(x)$ with 
$\sum_{i=1}^N\frac{\theta^2_i}{\lambda_i} <\infty$. 
Then the inner product $\langle\theta_1,\theta_2\rangle_{\mathcal H}$ between $\theta_1=\sum_{i=1}^N\theta_{1i}\phi_i(x)$,
and $\theta_2=\sum_{i=1}^N\theta_{2i}\phi_i(x)$
is of the form 
\begin{equation}
\langle\theta_1,\theta_2\rangle_{\mathcal H}=\sum_{i=1}^N\frac{\theta_{1i}\theta_{2i}}{\lambda_i}.
\label{eq:inner_product1}
\end{equation}
This induces the norm $\|\cdot\|_{\mathcal H}$, where 
$\|\theta\|^2_{\mathcal H}=\sum_{i=1}^N\frac{\theta^2_{i}}{\lambda_i}$. A smoothness condition on the space
is immediately imposed by requiring the norm to be finite -- the eigenvalues must decay sufficiently fast.

The Hilbert space defined above is a unique RKHS with respect to $\mathcal K$, 
in that it satisfies the following reproducing property:
\begin{equation}
\langle\theta,\mathcal K(\cdot,x)\rangle = \sum_{i=1}^N\frac{\theta_i\lambda_i\phi_i(x)}{\lambda_i} = \theta(x).
\label{eq:reproduce1}
\end{equation}
Further, the kernel satisfies the following:
\begin{equation}
\langle\mathcal K(x,\cdot),\mathcal K(x',\cdot)\rangle 
= \sum_{i=1}^N\frac{\lambda^2_i\phi_i(x)}{\lambda_i} = \mathcal K(x,x').
\label{eq:reproduce2}
\end{equation}

Now, with reference to (\ref{eq:eigen_Sigma_K}), observe that the square norm 
$\|\theta\|^2_{\mathcal H}=\sum_{i=1}^N\theta^2_i/\lambda_i$ and the quadratic form
$\btheta^T\mathcal\bK\btheta$ have the same form if the latter is expressed in terms of the eigenvectors of $\mathcal\bK$,
albeit the latter has $n$ terms, while the square norm has $N$ terms.

\subsection{Regularization}
\label{subsec:regularization}

The ill-posed-ness of inverse problems can be understood from the fact that for any given data set $\by_n$,
all functions that pass through the data set minimize any given measure of discrepancy $\mathbb D(\by_n,\btheta)$ 
between the data $\by_n$ and $\btheta$. To combat this, one considers minimization of the following regularized
functional:
\begin{equation}
R(\theta)=\mathbb D(\by_n,\btheta)+\frac{\tau}{2}\|\theta\|^2_{\mathcal H},
\label{eq:R_theta}
\end{equation}
where the second term, which is the regularizer, controls smoothness of the function and $\tau$ is the appropriate 
Lagrange multiplier.

The well-known representer theorem (see, for example, \ctn{Kimeldorf71}, \ctn{Finbarr86}, \ctn{Wahba90}, \ctn{Scholkopf02})
guarantees that each minimizer $\theta\in\mathcal H$ can be represented as
$\theta(x)=\sum_{i=1}^nc_i\mathcal K\left(x,x_i\right)$, where $\mathcal K$ is the corresponding reproducing kernel. 
If $\mathbb D\left(\by_n,\btheta\right)$ is convex, then there is a unique minimizer $\hat\theta$.

\subsection{Gaussian process modeling of the unknown function $\theta$}
\label{subsec:gp}

For simplicity, let us consider the model 
\begin{equation}
y_i=\theta(x_i)+\epsilon_i,
\label{eq:gp_model}
\end{equation}
for $i=1,\ldots,n$, where $\epsilon_i\stackrel{iid}{\sim}N(0,\sigma^2)$, where we assume $\sigma$
to be known for simplicity of illustration. Let $\theta(x)$ be modeled by a 
Gaussian process with mean function $\mu(x)$ and covariance kernel
$\mathcal K$ associated with the RKHS. In other words, for any $x\in\mathfrak X$, $E\left[\theta(x)\right]=\mu(x)$
and for any $x_1,x_2\in\mathfrak X$, $Cov\left(\theta(x_1),\theta(x_2)\right)=\mathcal K(x_1,x_2)$.

Assuming for convenience that $\mu(x)=0$ for all $x\in\mathfrak X$, it follows that the posterior 
distribution of $\theta(x^*)$ for any $x^*\in\mathfrak X$ %not included in the design set $\bx_n$
is given by 
\begin{equation}
\pi(\theta(x^*)|\by_n,\bx_n)\equiv N\left(\hat\mu(x^*),\hat\sigma^2(x^*)\right),
\label{eq:gp_posterior1}
\end{equation}
where, for any $x^*\in\mathfrak X$, 
\begin{align}
\hat\mu(x^*)&=\bs^T(x^*)\left(\mathcal \bK+\sigma^2\mathbb I_n\right)^{-1}\by_n;
\label{eq:gp_post_mean1}\\
\hat\sigma^2(x^*)&=\mathcal K(x^*,x^*)-\bs^T(x^*)\left(\mathcal bK+\sigma^2\mathbb I_n\right)^{-1}\bs(x^*),
\label{eq:gp_post_var1}
\end{align}
with $\bs(x^*)=\left(\mathcal K(x^*,x_1),\ldots,\mathcal K(x^*,x_n)\right)^T$.

Observe that the posterior mean admits the following representation:
\begin{equation}
\hat\mu(x^*) = \sum_{i=1}^n\tilde c_i\mathcal K(x^*,x_i),
\label{eq:representer1}
\end{equation}
where $\tilde c_i$ is the $i$-th element of $\left(\mathcal \bK+\sigma^2\mathbb I_n\right)^{-1}\by_n$.

In other words, the posterior mean of the Gaussian process based model is consistent with the representer
theorem.

\section{Regularization using differential operators and connection with Gaussian process}
\label{sec:diff_gp}

For $x=(x_1,\ldots,x_d)^T\in\mathbb R^d$, let 
\begin{equation}
\|\mathcal L^m\theta\|^2
=\int\sum_{j_1+\cdots+j_d=m}\left(\frac{\partial^m\theta(x)}{\partial x^{j_1}_1\cdots\partial x^{j_d}_d}\right)^2,
\label{eq:diff_op1}
\end{equation}
and
\begin{equation}
\|\mathcal P\theta\|^2=\sum_{m=0}^Mb_m\|\mathcal L^m\theta\|^2,
\label{eq:diff_op2}
\end{equation}
for some $M>0$, where the co-efficients $b_m\geq 0$. In particular, we assume for our purpose that $b_0>0$.
It is clear that $\|\mathcal P\theta\|^2$ is translation and rotation invariant. This norm penalizes $\theta$
in terms of its derivatives up to order $M$.

\subsection{Relation to RKHS}
\label{subsec:relation_rkhs}

It can be shown, using the fact that the complex exponentials $\exp(2\pi i s^Tx)$ are eigen functions
of the differential operator, that
\begin{equation}
\|\mathcal P\theta\|^2=\int\sum_{m=0}^Mb_m\left(4\pi^2s^Ts\right)^m\left|\tilde\theta(s)\right|^2ds,
\label{eq:diff_op3}
\end{equation}
where $\tilde\theta(s)$ is the Fourier transform of $\theta(s)$. Comparison of (\ref{eq:diff_op3}) with
(\ref{eq:inner_product1}) yields the power spectrum of the form $\left[\sum_{m=0}^Mb_m\left(4\pi^2s^Ts\right)^m\right]^{-1}$
which yields the following kernel by Fourier inversion:
\begin{equation}
\mathcal K(x,x')=\mathcal K(x-x')=\int\frac{\exp(2\pi i s^T(x-x'))}{\sum_{m=0}^Mb_m\left(4\pi^2s^Ts\right)^m}ds.
\label{eq:rk1}
\end{equation}

Calculus of variations can also be used to minimize $R(\theta)$ with respect to $\theta$, which yields (using
the Euler-Lagrange equation)
\begin{equation}
\theta(x)=\sum_{i=1}^nb_i\mathcal G(x-x_i),
\label{eq:el1}
\end{equation}
with
\begin{equation}
\sum_{i=1}^m(-1)^mb_m\nabla^m\mathcal G=\delta_{x-x'},
\label{eq:el2}
\end{equation}
where $\mathcal G$ is known as the Green's function. Using Fourier transform on (\ref{eq:el2}) it can be shown
that the Green's function is nothing but the kernel $\mathcal K$ given by (\ref{eq:rk1}). Moreover, it follows from
(\ref{eq:el2}) that $\sum_{i=1}^m(-1)^mb_m\nabla^m$ and $\mathcal K$ are inverses of each other.

Examples of kernels derived from differential operators are as follows. For $d=1$, setting $b_0=b^2$, $b_1=1$ and
$b_m=0$ for $m\geq 2$, one obtains $\mathcal K(x,x')=\mathcal K(x-x')=\frac{1}{2b}\exp\left(-b|x-x'|\right)$,
which is the covariance of the Ornstein-Uhlenbeck process. For general $d$ dimension, setting $b_m=b^{2m}/(m!2^m)$,
yields 
$\mathcal K(x,x')=\mathcal K(x-x')=\frac{1}{\left(2\pi b^2\right)^{d/2}}\exp\left[-\frac{1}{2b^2}(x-x')^T(x-x')\right]$.

Considering a grid $\bx_n$, note that 
\begin{equation}
\|\mathcal P\theta\|^2\approx \sum_{m=0}^Mb_m\left(D_m\btheta\right)^T\left(D_m\btheta\right)
=\btheta^T\left(\sum_{m=0}^MD^T_mD_m\right)\btheta,
\label{eq:approx1}
\end{equation}
where $D_m$ is a suitable finite-difference approximation of the differential
operator. Note that such finite-difference approximation has been explored in Section \ref{sec:linear_inverse_problem},
which we now investigate in a rigorous setting.
Also, since (\ref{eq:approx1}) is quadratic in $\btheta$, assuming a prior for $\btheta$, 
the logarithm of which has this form, and 
further assuming that $\log\left[\mathbb D(\by_n,\btheta)\right]$ is a log-likelihood quadratic in $\btheta$,
a Gaussian posterior results.

\subsection{Spline models and connection with Gaussian process}
\label{subsec:splines}

Let us consider the penalty function to be $\|\mathcal L^m\theta\|^2$. Then polynomials up to degree $m-1$
are not penalized and so, are in the null space of the regularization operator. In this case, it can be shown that
a minimizer of $R(\theta)$ is of the form
\begin{equation}
\theta(x)=\sum_{j=1}^kd_j\psi_j(x)+\sum_{i=1}^nc_iG(x,x_i),
\end{equation}
where $\{\psi_1,\ldots,\psi_k\}$ are polynomials that span the null space and the Green's function $G$ is given by
(see \ctn{Duchon77}, \ctn{Meinguet79})
\begin{equation}
G(x,x')=G(x-x')=\left\{\begin{array}{cc}c_{m,d}|x-x'|^{2m-d}\log |x-x'| & \mbox{if}~2m>d~\mbox{and}~d~\mbox{even}\\
c_{m,d}|x-x'|^{2m-d} & \mbox{otherwise}.\end{array}\right. ,
\end{equation}
where $c_{m,D}$ are constants (see \ctn{Wahba90} for the explicit form).

We now specialize the above arguments to the spline set-up. As before, let us consider
the model $y_i=\theta(x_i)+\epsilon_i$, where, for $i=1,\ldots,n$, $\epsilon_i\stackrel{iid}{\sim}N\left(0,\sigma^2\right)$.
For simplicity, we consider the one-dimensional set-up, and consider the cubic spline smoothing
problem that minimizes
\begin{equation}
R(\theta)=\sum_{i=1}^n(y_i-\theta(x_i))^2+\tau\int_0^1\left[\theta''(x)\right]^2dx,
\label{eq:spline1}
\end{equation}
where $0<x_1<\cdots<x_n<1$. The solution to this minimization problem is given by
\begin{equation}
\theta(x)=\sum_{j=0}^1d_jx^j+\sum_{i=1}^nc_i(x-x_i)^3_{+},
\label{eq:spline2}
\end{equation}
where, for any $x$, $(x)_+= x$ if $x>0$ and zero otherwise.

Following \ctn{Wahba78}, let us consider 
\begin{equation}
f(x)=\sum_{j=0}^1\beta_jx^j+\theta(x),
\label{eq:spline3}
\end{equation}
where $\bbeta=(\beta_0,\beta_1)^T\sim N\left(\bzero,\sigma^2_{\beta}\mathbb I_2\right)$, and
$\theta$ is a zero mean Gaussian process with covariance
\begin{equation}
\sigma^2_{\theta}\mathcal K(x,x')=\int_0^1(x-u)_{+}(x'-u)_{+}du=\sigma^2_\theta\left(\frac{|x-x'|v^2}{2}+\frac{v^3}{3}\right),
\label{eq:spline_cov1}
\end{equation}
where $v=\min\{x,x'\}$.

Taking $\sigma^2_{\beta}\rightarrow\infty$ makes the prior of $\bbeta$ vague, so that penalty on the polynomial
terms in the null space is effectively washed out.
It follows that 
\begin{equation}
E\left[\theta(x^*)|\by_n,\bx_n\right]=\bh(x^*)^T\hat\bbeta+\bs(x^*)^T\hat\bK^{-1}\left(\by_n-\bH^T\hat\bbeta\right),
\label{eq:spline_mean1}
\end{equation}
where, for any $x$, $\bh(x)=(1,x)^T$, $\bH=(\bh(x_1),\ldots,\bh(x_n))$, $\hat \bK$ is the covariance matrix
corresponding to $\sigma^2_\theta\mathcal K(x_i,x_j)+\sigma^2\delta_{ij}$, and
$\hat\bbeta=\left(\bH\hat\bK^{-1}\bH\right)^{-1}\bH\hat\bK^{-1}\by_n$.

Since the elements of $\bs(x^*)$ are piecewise cubic polynomials, it is easy to see that the posterior mean
(\ref{eq:spline_mean1}) is also a piecewise cubic polynomial. It is also clear that (\ref{eq:spline_mean1})
is a first order polynomial on $[0,x_1]$ and $[x_n,1]$.

\subsubsection{Connection with the $\ell$-fold integrated Wiener process}
\label{eq:l_fold}

\ctn{Shepp66} considered the $\ell$-fold integrated Wiener
process, for $\ell=0,1,2\ldots$, as follows:
\begin{equation}
W_{\ell}(x)=\int_0^1\frac{(x-u)^{\ell}_{+}}{\ell !}Z(u)du,
\label{eq:l_fold1}
\end{equation}
where $Z$ is a Gaussian white noise process with covariance $\delta(u-u')$.
As a special case, note that $W_0$ is the standard Wiener process. In our case, note that
\begin{equation}
\mathcal K(x,x')=Cov\left(W_1(x),W_1(x')\right).
\label{eq:cov1}
\end{equation}

The above ideas can be easily extended to the case of the regularizer $\int \left[f^{(m)}(x)\right]^2dx$, for $m\geq 1$
by replacing $(x-u)_{+}$ with $(x-u)^{m-1}_{+}/(m-1)!$ and letting $\bh(x)=\left(1,x,\ldots,x^{m-1}\right)^T$.

\section{The Bayesian approach to inverse problems in Hilbert spaces}
\label{sec:Bayesian_Hilbert}

We assume the following model
\begin{equation}
y=G(\theta)+\epsilon,
\label{eq:inv2}
\end{equation}
where $y$, $\theta$ and $\epsilon$ are in Banach or Hilbert spaces.

\begin{comment}
\subsection{ Notations and preliminaries }
%The Gaussian distribution with mean $\mu$ and variance $\sigma^2$ will be denoted as $\mathbf{N}(\mu, \sigma^2)$. 
%Let $\left( \mathbb{H}^{t}, <\cdot, \cdot>_{t} \right)_{t \in \Re}$ be Hilbert scale 
%(the detailed discussion of `Hilbert scale' can be found in  \ctn{hei} and also in \ctn{vol}).  
Let $\Gamma$ be positive definite trace-class linear operator with eigen system $(\lambda_{k}^{2}, \phi_{k})$. 
%(for definition of trace-class see Appendix 2 \eqref{tr}). 
This can be likened to trace of infinite-dimensional matrices 
with finite trace. Let us define $<u,v>_{t} := \left\langle \Gamma^{-\frac{t}{2}}u,  \Gamma^{-\frac{t}{2}}v \right \rangle$ 
with the norm $\Vert u\Vert_{t} := \Vert \Gamma^{-\frac{t}{2}}u  \Vert$. Define the ball with respect to 
$\Vert \cdot\Vert_{t}$ norm as $B_{r}^{t}(x) := \{ m : \Vert x-m \Vert_{t} < r \}$. For a given covariance 
$\Gamma$ let $\Vert \cdot\Vert_{\Gamma}$ denote the norm of Cameron-Martin space $\bar{\mathbb{B}}$ 
which is closure of Hilbert space $\mathbb{B}$ with respect to $<\cdot, \cdot>_{\Gamma}$. Similar to  \ctn{Vollmer13}, 
we define support of a measure $\mu$ in a metric space $(X, d)$  by 
$supp_{d}(\mu) = \left\lbrace  x \vert  \ \mu(B_{\delta}^{d}(x) > 0 \  \forall \  \delta >0 ) \right \rbrace$. 
%~~~~~~~~~~~~~~~~~~~~~~~~~

\end{comment}

\subsection{Bayes theorem for general inverse problems}
\label{subsec:Bayes_theorem}
We will consider the model stated by equation \eqref{eq:inv2}. Let $\mathcal Y$ and $\Theta$ denote the 
sample spaces for $y$ and $\theta$, respectively. Let us first assume that both are separable Banach spaces.
Assume $\mu_0$ to be the prior measure for $\theta$. Assuming well-defined joint distribution for 
$(y, \theta)$, let us denote the posterior of $\theta$ given $y$ as $\mu_y$. Let $\epsilon \sim Q_0$ where 
$Q_0$ %is a measure on $\mathbb{B}$ 
such that $\epsilon$ and $\theta$ are independent. %$\epsilon \perp X$. 
Let $Q_0$ be the distribution of $\epsilon$. 
%Let also $G \sim Q_{G}$ for some suitable measure $Q_{G}$. %For example, we can take $Q_G$ to be Gaussian process. 
Let us denote the conditional distribution of $y$ given $\theta$ by $Q_{\theta}$, obtained from a translation of 
$Q_{0}$ by $G(\theta)$. Assume that $Q_{\theta}\ll Q_0$. Thus, for some potential 
$\Phi: \Theta\times \mathcal Y \mapsto\mathbb R$,
\begin{equation}
\frac{dQ_{\theta}}{dQ_0}=\exp\left(-\Phi(\theta,y)\right).
\label{eq:rn1}
\end{equation}
Thus, for fixed $\theta$, $\Phi(\theta,\cdot):\mathcal Y\mapsto\mathbb R$ is measurable and
$E_{Q_0}\left[\exp\left(-\Phi(\theta,y)\right)\right]=1$. Note that $-\Phi(\cdot,y)$ is nothing but
the log-likelihood.

Let $\nu_0$ denote the product measure
\begin{equation}
\nu_0(d\theta,dy)=\mu_0(d\theta)Q_0(dy),
\label{eq:nu_0}
\end{equation}
and let us assume that $\Phi$ is $\nu_0$-measurable. Then $(\theta,y)\in\Theta\times\mathcal Y$ is distributed
according to the measure $\nu(d\theta,dy)=\mu_0(d\theta)Q_{\theta}(dy)$. It then also follows that
$\nu\ll\nu_0$, with
\begin{equation}
\frac{d\nu_{\theta}}{d\nu_0}(\theta,y)=\exp\left(-\Phi(\theta,y)\right).
\label{eq:rn2}
\end{equation}
Then we have the following statement of Bayes' theorem for general inverse problems:
\begin{theorem}[Bayes theorem for general inverse problems]
\label{theorem:bayes_theorem1}
Assume that $\Phi:\Theta\times\mathcal Y\mapsto\mathbb R$ is $\nu_0$-measurable and 
\begin{equation}
C=\int_{\Theta}\exp\left(-\Phi(\theta,y)\right)\mu_0(dy)>0,
\label{eq:positivity1}
\end{equation}
for $Q_0$-almost surely all $y$. Then the posterior of $\theta$ given $y$, which we denote by $\mu^y$,
exists under $\nu$. Also, $\mu^y\ll\mu_0$ and for all $y$ $\nu_0$-almost surely, 
\begin{equation}
\frac{d\mu^y_{\theta}}{d\mu_0}(\theta)=\frac{1}{C}\exp\left(-\Phi(\theta,y)\right).
\label{eq:bayes_theorem}
\end{equation}
\end{theorem}

Now assume that $\Theta$ and $\mathcal Y$ are Hilbert spaces. 
Suppose  $\epsilon \sim \mathbf{N}(0, \Gamma)$. Then the following theorem holds: 
%in \ctn{vol} and in \ctn{the}.
 
\begin{theorem}[\ctn{Vollmer13}]
\label{Ti1}
\begin{equation}
\label{hil1}
\dfrac{d\mu^{y}}{d\mu_{0}} \propto 
\exp \left( -\frac{1}{2} \Vert G(\theta)\Vert_{\Gamma}^{2} +\langle y, G(\theta)\rangle_{\Gamma} \right),
\end{equation}
\end{theorem}
where $\langle \cdot,\cdot\rangle_{\Gamma}=\langle\Gamma^{-1}\cdot,\cdot\rangle$, and
$\Vert\cdot\Vert_{\Gamma}$ is the norm induced by $\langle \cdot,\cdot\rangle_{\Gamma}$.

For the model $y_i=\theta(x_i)+\epsilon_i$ for $i=1,\ldots,n$, with $\epsilon_i\stackrel{iid}{\sim}N\left(0,\sigma^2\right)$,
the posterior is of the form
\begin{equation}
\dfrac{d\mu^{y}}{d\mu_{0}} \propto\exp\left(-\sum_{i=1}^n\frac{\left(y_i-\theta(x_i)\right)^2}{2\sigma^2}\right).
\end{equation}

\subsection{Connection with regularization methods}
\label{subsec:reg_con}
It is not immediately clear if the Bayesian approach in the Hilbert space setting has connection with
the deterministic regularization methods, but \ctn{Vollmer13} prove consistency of the posterior assuming
certain stability results which are used to prove convergence of regularization methods; see \ctn{Engl96}.

We next turn to inverse regression.

\section{Inverse regression}
\label{sec:inv_reg}

We first provide some examples of inverse regression, mostly based on \ctn{Avenhaus80}.
\subsection{Examples of inverse regression}
\label{subsec:examples}

\subsubsection{Example 1: Measurement of nuclear materials}
\label{subsubsec:example1}

Measurement of the amount of nuclear materials such as plutonium by direct chemical means is an extremely difficult exercise.
This motivates model-based methods. For instance, there are physical laws 
relating heat production or the number of neutrons
emitted (the dependent response variable $y$) to the amount of material present, 
the latter being the independent variable $x$. 
But any measurement instrument based on the physical laws first needs to be calibrated. In other words, 
%the more common vocabulary,
the unknown parameters of the model needs to be learned, using known inputs and outputs. 
However, the independent variables are usually subject to measurement errors, motivating a statistical model.
Thus, conditionally on $x$ and parameter(s) $\theta$, $y\sim P(\cdot|x,\theta)$, where $P(\cdot|x,\theta)$
denotes some appropriate probability model. Given $\by_n$ and $\bx_n$, and some specific $\tilde y$, the corresponding
$\tilde x$ needs to be predicted.

\subsubsection{Example 2: Estimation of family incomes}
\label{subsubsec:example2}

Suppose that it is of interest to estimate the family incomes in a certain city through public opinion poll.
Most of the population, however, will be unwilling to provide reliable answers to the questionnaires. One way
to extract relatively reliable figures is to consider some dependent variable, say, housing expenses ($y$), which
is supposed to strongly depend on family income ($x$); see \ctn{Muth60}, and such that the population is less
reluctant to divulge the correct figures related to $y$. From past survey data on $\bx_n$ and $\by_n$, and 
using current data from families who may provide reliable answers related to both $x$ and $y$, a statistical
model may be built, using which the unknown family incomes may be predicted, given their household incomes.

\subsubsection{Example 3: Missing variables}
\label{subsubsec:example3}

In regression problems where some of the covariate values $x_i$ are missing, they may
be estimated from the remaining data and the model. In this context, \ctn{Press75}
considered a simple linear regression problem in a Bayesian framework.
Under special assumptions about the error and prior distributions, they showed that an optimal procedure 
for estimating the linear parameters is to first estimate the missing $x_i$ from
an inverse regression based only on the complete data pairs.

\subsubsection{Example 4: Bioassay}
\label{subsubsec:example4}

It is usual to investigate the effects of substances ($y$) given
in several dosages on organisms ($x$) using bioassay methods. 
In this context it may be of interest to determine the dosage necessary to obtain some
interesting effect, making inverse regression relevant (see, for example, \ctn{Rasch73}).

\subsubsection{Example 5: Learning the Milky Way}
\label{subsubsec:astro}

The modelling of the Milky Way galaxy is an integral step in the study of galactic dynamics;
this is because knowledge of model parameters that define the Milky Way directly influences our
understanding of the evolution of our galaxy. Since the nature of the Galaxy's phase space, in the
neighbourhood of the Sun, is affected by distinct Milky Way features, measurements of phase space
coordinates of individual stars that live in this neighbourhood of the Sun, will bear information
about the influence of such features. Then, inversion of such measurements can help us learn the
parameters that describe such Milky Way features. In this regard, learning about the location of the
Sun with respect to the center of the galaxy, given the two-component velocities of the stars in the vicinity of the Sun, 
is an important problem. 
For $k$ such stars, \ctn{Chakrabarty15} model the $k\times 2$-dimensional velocity matrix $\bV$ 
as a function of the galactocentric location ($\bS$) of the Sun, denoted by $\bV=\bxi(\bS)$. 
For a given observed value $\bV^*$ of $\bV$, it is then of interest to obtain the corresponding $\bS^*$.
Since $\bxi$ is unknown, \ctn{Chakrabarty15} model $\bxi$ as a matrix-variate Gaussian process, and consider
the Bayesian approach to learning about $\bS^*$, given data $\{(\bS_i,\bV_i):i=1,\ldots,n\}$ simulated
from established astrophysical models, and the observed velocity matrix $\bV^*$.

We now provide a brief overview of of the methods of inverse linear regression, 
which is the most popular among inverse regression problems. Our discussion is generally based on 
\ctn{Hoadley70} and \ctn{Avenhaus80}.

\subsection{Inverse linear regression}
\label{subsec:inv_linear_reg}

Let us consider the following simple linear regression model: for $i=1,\ldots,n$,
\begin{equation}
y_i=\alpha+\beta x_i+\sigma\epsilon_i,
\label{eq:lm1}
\end{equation}
where $\epsilon_i\stackrel{iid}{\sim}N(0,1)$.

For simplicity, let us consider a single unknown $\tilde x$, associated with a further set of $m$ responses 
$\{\tilde y_1,\dots,\tilde y_m\}$,
related by
\begin{equation}
\tilde y_i=\alpha+\beta \tilde x+\tau\tilde\epsilon_i,
\label{eq:lm2}
\end{equation}
for $i=1,\ldots,m$, where $\tilde\epsilon_i\stackrel{iid}{\sim}N(0,1)$ and are independent of the $\epsilon_i$'s associated
with (\ref{eq:lm1}).

The interest in the above problem is inference regarding the unknown $x$. 
Based on (\ref{eq:lm1}), first least squares estimates of $\alpha$ and $\beta$ are obtained as 
\begin{align}
\hat\beta &=\frac{\sum_{i=1}^n(y_i-\bar y)(x_i-\bar x)}{\sum_{i=1}^n(x_i-\bar x)^2};
\label{eq:hat_beta}\\
\hat\alpha &=\bar y-\hat\beta\bar x,
\label{eq:hat_alpha}
\end{align}
where $\bar y=\sum_{i=1}^ny_i/n$ and $\bar y=\sum_{i=1}^nx_i/n$. Then, letting
$\bar {\tilde y}=\sum_{i=1}^n\tilde y_i/n$, a `classical' estimator of $x$ is given by
\begin{equation}
\hat x_C=\frac{\bar {\tilde y}-\hat\alpha}{\hat\beta},
\label{eq:classical1}
\end{equation}
which is also the maximum likelihood estimator for the likelihood associated with (\ref{eq:lm1}) and (\ref{eq:lm2}),
assuming known $\sigma$ and $\tau$. 
However, %$E\left(\hat x_C|\alpha,\beta,\sigma,\tau,x\right)$ doe snot exist and
\begin{equation}
E\left[\left(\hat x_C-x\right)^2|\alpha,\beta,\sigma,\tau,x\right]=\infty,
\label{eq:mse_inf1}
\end{equation}
which prompted \ctn{Krutchkoff67} to propose the following `inverse' estimator:
\begin{equation}
\hat x_I=\hat\gamma+\hat\delta\bar{\tilde y},
\label{eq:x_I}
\end{equation}
where
\begin{align}
\hat\delta &=\frac{\sum_{i=1}^n(y_i-\bar y)(x_i-\bar x)}{\sum_{i=1}^n(y_i-\bar y)^2};
\label{eq:hat_delta}\\
\hat\gamma &=\bar x-\hat\delta\bar y,
\label{eq:hat_gamma}
\end{align}
are the least squares estimators of the slope and intercept when the $x_i$ are regressed on the $y_i$.
It can be shown that the mean square error of this inverse estimator is finite. However,
\ctn{Williams69} showed that if $\sigma^2=\tau^2$ and if the sign of $\beta$ is known, 
then the unique unbiased estimator of $x$ has infinite variance.
Williams advocated the use of confidence limits instead of point estimators.
%which should provide what
%is required for inverse linear regression. Hence the two papers of Perng
%and Tong (1974 and 1977) could meet his approval. They treated the problem
%of the allocation of n and m for the interval estimation of
%X
%so that the
%probability of coverage is maximized when the total number of observations
%n+m is fixed and is large.

\ctn{Hoadley70} derive confidence limits setting $\sigma=\tau$ and assuming without loss of generality that $\sum_{i=1}^nx_i=0$.
Under these assumptions, the maximum likelihood estimators of $\sigma^2$ with $\bx_n$ and $\by_n$ only,
$\tilde\by_n=(\tilde y_1,\ldots,\tilde y_n)^T$ only, and with the entire available data set are, respectively,
\begin{align}
\hat\sigma^2_1&=\frac{1}{n-2}\sum_{i=1}^n\left(y_i-\hat\alpha-\hat\beta x_i\right)^2;\label{eq:sig1}\\
\hat\sigma^2_2&=\frac{1}{m-1}\sum_{i=1}^n\left(\tilde y_i-\bar{\tilde y}\right)^2;\label{eq:sig2}\\
\hat\sigma^2&=\frac{1}{n-2+m-1}\left[(n-2)\sigma^2_1+(m-1)\sigma^2_2\right].\label{eq:sig3}
\end{align}
Now consider the $F$-statistic $F=\frac{n\hat\beta^2}{\hat\sigma^2}$ for testing the hypothesis $\beta=0$.
Note that under the null hypothesis this statistic has the $F$ distribution with $1$ and $n+m$ degrees of freedom. 
For $m=1$, $$\hat\beta\left(\hat x_C-x\right)\sqrt{\frac{n}{\sigma^2(n+1+x^2)}}$$ has a $t$ distribution with $n-2$ degrees
of freedom. Letting $F_{\alpha;1,\nu}$ denote the upper $\alpha$ point of the $F$ distribution with $1$ and $\nu$
degrees of freedom, a confidence set $S$ can be derived as follows:
\begin{equation}
S=\left\{\begin{array}{ccc} \{x:x_L\leq x\leq x_U\} & \mbox{if} & F>F_{\alpha;1,n-2};\\
\{x:x\leq x_L\}\cup\{x\geq x_U\} & \mbox{if} & \frac{n+1}{n+1+\hat x^2_C}F_{\alpha;1,n-2}\leq F< F_{\alpha;1,n-2};\\
(-\infty,\infty) & \mbox{if} & F<\frac{n+1}{n+1+\hat x^2_C}F_{\alpha;1,n-2},\end{array}\right.
%\label{eq:cl1}
\end{equation}
where $x_L$ and $x_U$ are given by
\begin{equation*}
\frac{F\hat x_C}{F-F_{\alpha;1,n-1}}\pm 
\frac{\left\{F_{\alpha;1,n-2}\left[(n+1)\left(F-F_{\alpha;1,n-2}\right)+F\hat x^2_C\right]\right\}^{\frac{1}{2}}}
{F-F_{\alpha;1,n-2}}.
\end{equation*}
Hence, if $F<\frac{n+1}{n+1+\hat x^2_C}F_{\alpha;1,n-2}$, then the associated confidence interval is 
$S=(-\infty,\infty)$, which is of course useless.

\ctn{Hoadley70} present a Bayesian analysis of this problem, presented below in the form of the following two theorems.
\begin{theorem}[\ctn{Hoadley70}]
\label{theorem:Hoadley1}
Assume that $\sigma=\tau$, and let $x$ be independent of $(\alpha,\beta,\sigma^2)$ {\it a priori}. With any prior
$\pi(x)$ on $x$ and the prior
\begin{equation*}
\pi(\alpha,\beta,\sigma^2)\propto\frac{1}{\sigma^2}
\end{equation*}
on $(\alpha,\beta,\sigma^2)$, the posterior density of $x$ given by
\begin{equation*}
\pi(x|\by_n,\bx_n,\tilde\by_n)\propto\pi(x)L(x),
\end{equation*}
where
\begin{equation*}
L(x)=\frac{\left(1+\frac{n}{m}+x^2\right)^{\frac{m+n-3}{2}}}
{\left[1+\frac{n}{m}+R\hat x^2_C+\left(\frac{F}{m+n-3}+1\right)\left(x-R\hat x_C\right)^2\right]^{\frac{m+n-2}{2}}},
\end{equation*}
where $$R=\frac{F}{F+m+n-3}.$$
\end{theorem}

For $m=1$, \ctn{Hoadley70} present the following result characterizing the inverse estimator $\hat x_I$:
\begin{theorem}[\ctn{Hoadley70}]
\label{theorem:Hoadley2}
Consider the following informative prior on $x$:
$$x=t_{n-3}\frac{n+1}{n-3},$$
where $t_{\nu}$ denotes the $t$ distribution with $\nu$ degrees of freedom. Then the posterior distribution of $x$
given $\by_n$, $\bx_n$ and $\tilde\by_n$ has the same distribution as
$$\hat x_I+t_{n-2}\sqrt{\frac{n+1+\frac{\hat x^2_I}{R}}{F+n-2}}.$$
\end{theorem}
In particular, it follows from Theorem \ref{theorem:Hoadley2} that the posterior mean of $x$ is $\hat x_I$ when $m=1$.
In other words, the inverse estimator $\hat x_I$ is Bayes with respect to the squared error loss and a particular
informative prior distribution for $x$.

Since the goal of \ctn{Hoadley70} was to provide a theoretical justification
of the inverse estimator, he had to choose a somewhat unusual prior so that it leads to $\hat x_I$ as the posterior
mean. In general it is not necessary to confine ourselves to any specific prior for Bayesian
analysis of inverse regression. It is also clear that the Bayesian framework is appropriate 
for any inverse regression problem, not just linear inverse regression; indeed, the palaeoclimate
reconstruction problem (\ctn{Haslett06}) and the Milky Way problem (\ctn{Chakrabarty15}) are examples
of very highly non-linear inverse regression problems.
%\ctn{Avenhaus80} propose two other estimators of $x$; all the estimators including the classical and
%the inverse estimator are linear in $\bar{\tilde y}$, but .

\subsection{Connection between inverse regression problems and traditional inverse problems}
\label{subsec:inverse_inverse_reg}
Note that the class of inverse regression problems includes the class of traditional inverse problems.
The Milky Way problem is an example where learning the unknown, matrix-variate function $\bxi$ (inverse problem)
was required, even though learning about $\bS$, the galactocentric location of the sun (inverse regression problem) 
was the primary goal. 
The Bayesian approach allowed learning both $\bS$ and $\bxi$ simultaneously and coherently.

In the palaeoclimate models proposed in \ctn{Haslett06}, \ctn{Bhatta06} and \ctn{Sabya13}, although
species assemblages are modeled conditionally on climate variables, the functional relationship between
species and climate are not even approximately known. In all these works, it is of interest to learn about
the functional relationship as well as to predict the unobserved climate values, the latter being the main aim.
Again, the Bayesian approach facilitated appropriate learning of both the unknown quantities.

\subsection{Consistency of inverse regression problems}
\label{subsec:consistency_inverse_regression}
In the above linear inverse regression, notice that if $\tau>0$, then the variance of the estimator of $x$
can not tend to zero, even as the data size tends to infinity. This shows that no estimator of $x$
can be consistent. The same argument applies even to Bayesian approaches; for any sensible prior on $x$ that
does not give point mass to the true value of $x$, the posterior of $x$ will not converge to the point mass
at the true value of $x$ as the data size increases indefinitely. The arguments remain valid for any inverse
regression problem where the response variable $y$ probabilistically depends upon the independent variable $x$.
Not only in inverse regression problems, even in forward regression problems where the interest is in prediction
of $y$ given $x$, any estimate of $y$ or any posterior predictive distribution $y$ will be inconsistent.

\begin{figure}
\begin{center}
\includegraphics[width=9cm,height=8cm]{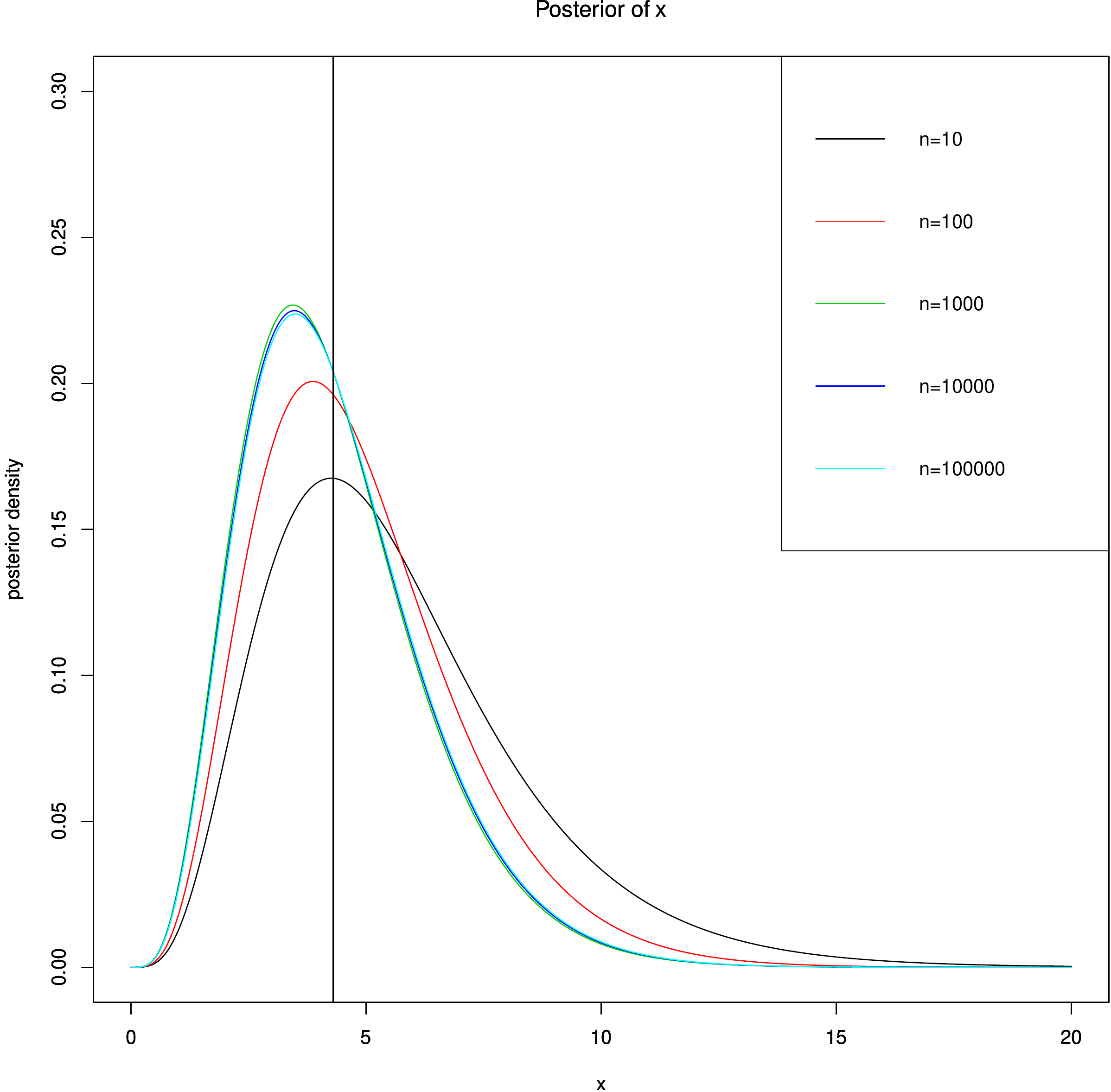}
\end{center}
\caption{Demonstration of posterior inconsistency in inverse regression problems. The vertical line denotes the true value.}
\label{fig:xval_inverse}
\end{figure}

To give an example of inconsistency in non-linear and non-normal inverse problem, consider the following set-up:
$y_i\stackrel{iid}{\sim}\mbox{Poisson}\left(\theta x_i\right)$, for $i=1,\ldots,n$, where $\theta>0$ and $x_i>0$
for each $i$. Let us consider the prior $\pi(\theta)\equiv 1$ for all $\theta>0$. For some $i^*\in\{1,\ldots,n\}$
let us assume the leave-one-out cross-validation set-up in that we wish to learn $x=x_{i^*}$ assuming it is unknown,
from the rest of the data. Putting the prior $\pi(x)\equiv 1$ for $x>0$, the posterior of $x$ is given by 
(see \ctn{Bhattacharya07}, \ctn{Bhattacharya13})
\begin{equation}
\pi(x|\bx_n\backslash x_i,\by_n)\propto \frac{x^{y_i}}{(x+\sum_{j\neq i}x_j)^{(\sum_{j=1}^ny_j+1)}}.
\label{eq:xval}
\end{equation}
Figure \ref{fig:xval_inverse} displays the posterior of $x$ when $i^*=10$, for increasing sample size. Observe that the variance
of the posterior does not decrease even with sample size as large as $100,000$, clearly demonstrating inconsistency.
Hence, special, innovative priors are necessary for consistency in such cases.

\section{Conclusion}
\label{sec:conclusion}

In this review article, we have clarified the similarities and dissimilarities between the traditional inverse
problems and the inverse regression problems. In particular, we have argued that only the latter class of problems
qualify as authentic inverse problems in they have significantly different goals compared to the corresponding
forward problems. Moreover, they include the traditional inverse problems on learning unknown functions as a special case, 
as exemplified by our palaeoclimate and Milky Way examples. We advocate the Bayesian paradigm for both classes of problems,
not only because of its inherent flexibility, coherency and posterior uncertainty quantification, but also because
the prior acts as a natural penalty which is very important to regularize the so-called ill-posed inverse problems.
The well-known Tikhonov regularizer is just a special case from this perspective. 

It is important to remark that the literature on inverse function learning problems and inverse regression problems 
is still very young and a lot of research is necessary to develop the fields. Specifically, there is hardly
any well-developed, consistent model adequacy test or model comparison methodology in either of the two fields, 
although \ctn{Djafari00} deal with some specific inverse problems in this context, and \ctn{Bhattacharya13}
propose a test for model adequacy in the case of inverse regression problems. Moreover, as we have demonstrated,
inverse regression problems are inconsistent in general.
The general development in these respects will be provided in the PhD thesis of the first author. 

\pagebreak

\section*{References}
\bibliographystyle{ECA_jasa}
\bibliography{irmcmc}

\end{document}